\documentclass[letterpaper,aps,prb,showpacs,intlimits,amsmath,amssymb,twocolumn,superscriptaddress,floatfix]{revtex4}

\usepackage{graphicx}
\usepackage{bm}

\begin{document}

\title{Resonant Raman scattering effects in a nesting driven charge-density-wave insulator: exact analysis of the spinless Falicov-Kimball model with dynamical mean-field theory}

\author{O.~P.~Matveev}
\affiliation{Institute for Condensed Matter Physics of the
National Academy of Sciences of Ukraine, Lviv, 79011 Ukraine}

\author{A.~M.~Shvaika}
\affiliation{Institute for Condensed Matter Physics of the
National Academy of Sciences of Ukraine, Lviv, 79011 Ukraine}

\author{J.~K.~Freericks}
\affiliation{Department of Physics, Georgetown University,
Washington, DC 20057, U.S.A.}

\begin{abstract}
We calculate the total
electronic Raman scattering spectrum for a system with a charge density wave on an infinite-dimensional
hypercubic lattice. The problem is solved exactly for the
spinless Falicov-Kimball model with dynamical mean-field theory. We include the nonresonant, mixed, and resonant contributions in three common experimental polarizations, and analyze the response functions for representative values of the energy of the incident
photons. The complicated scattering response can be understood from the significant temperature dependence of the many-body density of states, and includes a huge enhancement for photon frequencies near the charge-density-wave gap energy.
\end{abstract}


\maketitle

\section{Introduction}

Inelastic light scattering is a powerful probe of the charge fluctuations in a strongly correlated material~\cite{Raman_review}.  By using polarizers on the incident and the reflected light, one can examine different symmetry channels for the charge excitations, and how easily they can scatter light.  Using inelastic light scattering, one can learn about the symmetry of underlying order, such as the d-wave superconductivity in the high temperature superconductors. Here, we will focus on the effects of static charge-density-wave (CDW) order on the inelastic light scattering of a strongly correlated material. 

The field of inelastic light scattering has been increasing in interest.  When x-rays are used for the light source, one can examine resonant inelastic x-ray scattering, where both energy and momentum are exchanged between the light and the charge excitations of the solid.  Here, we focus on the zero momentum limit, where only energy is exchanged, because we will be using optical light.  Hence we will be examining resonant effects in electronic Raman scattering.  The dynamical mean-field theory (DMFT) approach to this problem was completed a few years ago~\cite{sh_v_f_d1,sh_v_f_d2,sh_v_f_d3,sh_v_f_d4} in the normal state.  One of the interesting results from that work was that one could see a joint resonance of low-energy features, with higher-energy features when the photon energy was on the order of the interaction strength $U$ between the electrons. When one has charge-density-wave order, there are two additional complications that arise: (i) the density of states (DOS) has significant temperature dependence below $T_c$, where excitations with energies smaller than the gap energy will be depleted as $T\rightarrow 0$, and (ii) the DOS develops sharp, singular peaks as $T\rightarrow 0$ that arise at the gap edge.  One would hence expect the Raman response to have much more temperature dependence than what was seen in the normal state and to have more striking resonant effects because of the sharp peaks which develop due to a pile-up of the DOS at the gap edge.  Indeed, the nonresonant Raman response, in the CDW phase, shows dramatic effects due to the singularity in the DOS in some of the symmetry channels~\cite{MSF2}.

CDW order is also interesting because there are a number of strongly correlated materials that display this behavior.  The most prevalent class of such materials are the transition metal di- and trichalchogenides, which display either quasi one dimensional (NbSe$_3$) or quasi two dimensional (TaSe$_2$ or TbTe$_3$) CDW order~\cite{NbSe3,TaSe2,TbTe3}. In addition, there are known three-dimensional systems like BaBiO$_3$ and Ba$_{1-x}$K$_{x}$BiO$_3$ which display charge-density-wave order via nesting on a bipartite lattice at half filling~\cite{cdw_exp}.  This latter example is particularly relevant to our work, since the DMFT is more accurate as the dimensionality increases. One of the longstanding questions in the field is the question of whether the order is driven electronically, with a lattice instability following the electronic instability, or {\it vice versa}.  We won't have any direct answers to that question in this work, since we are not examining time-resolved phenomena, but we will note that experimental light scattering work has already examined the phonon softening phenomena that is associated with the lattice distortion~\cite{cooper}.  Here we focus on electronic effects, which would be the obvious next generation of experimental probes on these systems.  

We will be varying the photon energy over a wide range of different values.  We will see the most remarkable resonant effects when the photon energy is equal to the gap energy, as one might naively expect.  For many CDW systems, this gap energy is at most a few hundred meV, which is much below the optical photon energies. Hence, the experimentally most relevant results will rely on examining joint resonant effects, like what was observed in the normal state in previous calculations. But we also will focus some attention on the most dramatic resonant effects under the hope that such CDW systems, made from strongly correlated electronic systems, might be found in the future, and that they can be studied with electronic Raman scattering.

We use the Falicov-Kimball model in our analysis because it is one of the simplest models~\cite{falicov_kimball} which possesses static CDW ordering and has an exact solution within DMFT~\cite{brandt_mielsch1} (for a review see Ref.~\onlinecite{freericks_review}).  In particular, the irreducible charge vertex is known exactly, and that is needed to examine the charge screening effects.
Our work also extends recent results on transport, optical conductivity, and nonresonant x-ray scattering in CDW systems~\cite{krishnamurthy, MSF1, MSF2} to the realm of resonant inelastic light scattering. A brief report on resonant Raman scattering has also appeared as a conference proceeding~\cite{MSF3}.

The organization of this paper is as follows:  in section II, we introduce the model and briefly review the dynamical mean-field theory approach in the ordered phase; in section III, we describe the general formalism for inelastic light scattering; in section IV, we focus on the detailed formulas for the mixed and resonant contributions to Raman scattering; in section V, we present our numerical results and we analyze the Raman scattering response for two different cases; and in section VI, we present our conclusions.

\section{Ordered phase dynamical mean-field theory}

Historically, the Falicov-Kimball model~\cite{falicov_kimball} was introduced in 1969 to describe metal-insulator transitions in rare-earth compounds and transition-metal oxides involving a simplified two-band model with localized heavy electrons and itinerant light electrons which hop between sites. The
mobile electrons hop to neighboring sites with a hopping integral $-t$ and they interact with the localized particles at the same site with the Coulomb energy $U$. The mobile electron creation (annihilation) operator at site $i$ is denoted by $\hat d_i^\dagger$ ($\hat d_i^{}$) and the local electron creation (annihilation) operator at site $i$ is $\hat f_i^\dagger$ ($\hat f_i^{}$). We perform our calculations at half-filling because, in this case, there is an insulating CDW phase at low temperature for all values of $U$. The explicit formula for the Hamiltonian appears in Eqs.~(\ref{H_FK}--\ref{H_loc}).

An algorithm to determine the (period-two) ordered-phase Green functions (within DMFT) was developed by Brandt and
Mielsch~\cite{brandt_mielsch2} shortly after Metzner and Vollhardt introduced the idea of the many-body problem simplification in large dimensions~\cite{metzner_vollhardt}. The CDW order parameter displays anomalous behavior at weak coupling~\cite{vandongen,chen_freericks}, and higher-period ordered phases are possible, and have been examined on the Bethe lattice~\cite{freericks_swiss}. In previous works~\cite{krishnamurthy,MSF1,MSF2}, the transport properties and nonresonant inelastic light and x-ray  scattering were examined in the commensurate CDW phase. A detailed description of the DMFT solution for the CDW phase of the Falicov-Kimball model has also appeared in our previous papers~\cite{MSF1,MSF2}, so we restrict ourselves to a brief summary in order to establish our notation.

We work on an infinite-dimensional hypercubic lattice with nearest neighbor hopping.  This lattice is bipartite, implying that it can be divided into two sublattices, denoted $A$ and $B$, with the hopping being nonzero only between the different sublattices.  In this case, the Falicov-Kimball model has particle-hole symmetry, and the noninteracting Fermi surface is nested at half filling with an ordering wavevector at the zone boundary along the diagonal, which implies the CDW order will lie on the sublattice structure, with the density of the light and of the heavy electrons being uniform on each sublattice, but different on the different sublattices. This difference in electron filling serves as the order parameter for the CDW phase. Keeping this in the mind, we introduce sublattice indices into the Falicov-Kimball model Hamiltonian
\begin{equation}\label{H_FK}
  \hat{H}=\sum_{ia}\hat{H}_{i}^{a}-
  \sum_{ijab}t_{ij}^{ab}\hat{d}_{ia}^{\dag}\hat{d}_{jb}^{},
\end{equation}
where $i$ and $a=A$ or $B$ are the site and sublattice indices, respectively, and $t_{ij}^{ab}$ is the hopping matrix, which is nonzero only between different sublattices ($t_{ij}^{AA}=t_{ij}^{BB}=0$). The local part of the Hamiltonian is equal to
\begin{equation}\label{H_loc}
  \hat{H}_{i}^{a}=U\hat{n}_{id}^{a}\hat{n}_{if}^{a}-
  \mu_{d}^{a}\hat{n}_{id}^{a}-\mu_{f}^{a}\hat{n}_{if}^{a};
\end{equation}
with the number operators of the itinerant and localized electrons given by $\hat n_{id}=\hat d_i^\dagger \hat d_i^{}$ and $\hat n_{if}=\hat f_i^\dagger\hat f_i^{}$, respectively. For computational convenience, we have introduced different chemical potentials for different sublattices, which allows us to work with a fixed order parameter, rather than iterating the DMFT equations to determine the order parameter (which is subject to critical slowing down near $T_c$). The system achieves its equilibrium state when the chemical potential is uniform throughout the lattice ($\mu^A_d=\mu^B_d$ and $\mu^A_f=\mu^B_f$).

The first step of the DMFT approach is to scale the hopping matrix element as $-t=-t^*/2\sqrt{D}$ (we use $t^*=1$ as the unit of energy) and then take the limit of infinite dimensions $D\to\infty$.~\cite{metzner_vollhardt} The self-energy is then local:
\begin{equation}\label{Sigma_inf}
  \Sigma_{ij}^{ab}(\omega)=\Sigma_{i}^{a}(\omega)\delta_{ij}\delta_{ab},
\end{equation}
and in the case of two sublattices has two values $\Sigma^{A}(\omega)$ and $\Sigma^{B}(\omega)$. As a result, the DMFT equations become matrix equations for the CDW phase.  Hence, we can write the solution of the Dyson equation (in momentum space) in a matrix form
\begin{equation}\label{Dyson}
  \mathbf{G}_{\bm k}(\omega)=\left[\mathbf{z}(\omega)-\mathbf{t}_{\bm k}\right]^{-1},
\end{equation}
where the irreducible part $\mathbf{z}(\omega)$ and the hopping term $\mathbf{t}_{\bm k}$ are represented by the following $2\times2$ matrices:
\begin{align}\label{z-t}
  \mathbf{z}(\omega)&=\left ( \begin{array}{cccc}
  \omega+\mu^{A}_{d}-\Sigma^{A}(\omega) & 0  \\
  0 & \omega+\mu^{B}_{d}-\Sigma^{B}(\omega) \\
  \end{array}\right ),
  \\
  \mathbf{t}_{\bm k}&=\left (\begin{array}{cccc}
  0 & \epsilon_{\bm k}  \\
  \epsilon_{\bm k} & 0 \\
  \end{array}\right ),
  \nonumber
\end{align}
with the band structure $\epsilon_{\bm k}$ satisfying $\epsilon_{\bm k}=-t^*\lim_{D\rightarrow\infty}\sum_{i=1}^D\cos{\bm k}_i/\sqrt{D}$.
Then we can represent the local Green's function on sublattice $a$
\begin{equation}\label{G_loc}
  G^{aa}(\omega)=\frac1N\sum_{\bm k} G_{\bm k}^{aa}(\omega),
\end{equation}
in terms of the local dynamical mean field $\lambda^a(\omega)$, via
\begin{equation}\label{G_CPA}
  G^{aa}(\omega)=\frac{1}{\omega+\mu^a_d-\Sigma^a(\omega)-\lambda^a(\omega)}.
\end{equation}
Finally, we close the system of equations for $\Sigma^a(\omega)$ and $\lambda^a(\omega)$ by finding the local Green's function from the solution of an impurity problem in the dynamical mean field $\lambda^a(\omega)$. For the Falicov-Kimball model such a problem can be solved exactly  and the result is equal to
\begin{equation}\label{G_FKM}
  G^{aa}(\omega)=\frac{1-n_f^a}{\omega+\mu^a_d-\lambda^a(\omega)}
  + \frac{n_f^a}{\omega+\mu^a_d-U-\lambda^a(\omega)},
\end{equation}
where $n_f^a$ is the average concentration of the localized electrons on the sublattice $a$. In the CDW phase, the total concentration of localized electrons is fixed $n_f^A+n_f^B=\mathrm{const.}$ and the difference of the concentrations on each sublattice $\Delta n_f=n_f^A-n_f^B$ is the order parameter of the CDW phase and is defined from the equilibrium condition on the sublattice chemical potentials: $\mu_{f}^{A}-\mu_{f}^{B}=0$.

Numerical solutions of these equations are given in Ref.~\onlinecite{MSF1} where the evolution of the
DOS in the CDW-ordered phase is shown. At $T=0$, a real gap develops of magnitude $U$ with square root singularities at the band edges (even on the hypercubic lattice which has infinite tails to the DOS in the normal state). As the temperature increases, the system develops substantial subgap DOS which are thermally activated within the ordered phase.  Additional plots of the DOS can be found in Ref.~\onlinecite{MSF1}. Note that the singular behavior occurs for one of the ``inner'' band edges on each sublattice, and that the subgap states develop very rapidly as the temperature rises and completely fill in the CDW gap at the critical temperature $T_c$.

\section{Formalism for inelastic light scattering}

The interaction of a weak external transverse electromagnetic field \textbf{A} with an electronic system with nearest-neighbor hopping is described by the Hamiltonian~\cite{shastry_shraiman1,shastry_shraiman2}:
\begin{align}\label{H_int}
  H_{\textrm{int}}&=-\frac{e}{\hbar c}
  \sum_{\bm k}{\bm j(\bm k)\cdot {\bm A}(-\bm k)} \\
  &+\frac{e^{2}}{2\hbar^{2}c^{2}}
  \sum_{\bm k \bm k'}\sum_{\alpha\beta}{A_{\alpha}(-\bm k)\gamma_{\alpha,\beta}(\bm k+\bm k')A_{\beta}(-
  \bm k')},
  \nonumber
\end{align}
where the current operator and stress tensor for itinerant electrons are equal to
\begin{align}\label{eq:current}
  j_{\alpha}(\bm q)&=\sum_{ab\bm k}{
 \frac{\partial t_{ab}(\bm k)}{\partial
  k_{\alpha}}\hat{d}_{a}^{\dag}(\bm k+\bm q/2)\hat{d}_{b}(\bm k-\bm q/2)}
\end{align}
and
\begin{align}\label{eq:stress}
  \gamma_{\alpha,\beta}(\bm q)=\sum_{ab\bm k}
  \frac{\partial^{2} t_{ab}(\bm k)}{\partial k_{\alpha}
  \partial k_{\beta}}\hat{d}_{a}^{\dag}(\bm k+\bm q/2)\hat{d}_{b}(\bm k-\bm q/2),
\end{align}
respectively. Here $t_{ab}(\bm k)$ are the components of the $2\times 2$ hopping matrix in Eq.~(\ref{z-t}). The general formula for the
inelastic light scattering cross section
\begin{align}\label{eq11}
  R(\bm q,\Omega)&=2\pi\sum_{i,f}
 \frac{e^{-\beta\varepsilon_{i}}}{\mathcal Z}
  \delta(\varepsilon_{f}-\varepsilon_{i}-\Omega) \\
  &\times \left|\sum_{\alpha\beta}g(\bm k_{i})g(\bm k_{f})e_{\alpha}^{i}e_{\beta}^{f}
  \left\langle f\left|\hat{M}^{\alpha\beta}(q)\right|i\right\rangle \right|^{2}.
  \nonumber
\end{align}
is expressed~\cite{shastry_shraiman1,shastry_shraiman2} through the square of the scattering operator
\begin{align}\label{eq12}
  \left\langle f\left|\hat{M}^{\alpha\beta}(\bm q)\right|i\right\rangle
  &=\left\langle f\left|\gamma_{\alpha,\beta}(\bm
  q)\right|i\right\rangle \\
  &+\sum_{l}\Biggl(\frac{\left\langle f\left|j_{\beta}(\bm k_{f})
  \right|l\right\rangle  \left\langle l\left|j_{\alpha}(-\bm k_{i})
  \right|i\right\rangle  }{\varepsilon_{l}-\varepsilon_{i}-\omega_{i}}
  \nonumber \\
  &+\frac{\left\langle f\left|j_{\alpha}(-\bm k_{i})
  \right|l\right\rangle  \left\langle l\left|j_{\beta}(\bm k_{f})
  \right|i\right\rangle }{\varepsilon_{l}-\varepsilon_{i}+\omega_{f}}
  \Biggr)
  \nonumber
\end{align}
which contains both nonresonant and resonant contributions. Here $\Omega=\omega_{i}-\omega_{f}$ and $\bm q=\bm k_{i}-\bm k_{f}$ are the transferred energy and momentum of the photons, respectively, ${\bm e}^{i(f)}$ is the polarization of the initial (final) states of the photons and $\varepsilon_{i(f)}$ denotes the electronic energies for the initial $i$ and final $f$ electronic eigenstates. The quantity $g(\bm q)=(hc^{2}/V\omega_{\bm q})^{1/2}$ is called the ``scattering strength'' with $\omega_{\bm q}=c|\bm q|$, and ${\mathcal Z}$ is the partition function for the electronic system. The nonresonant part of the scattering operator $\hat{M}(\bm q)$ is constructed from the stress tensor and the resonant one is constructed from the square of the current operators. After substituting the expression for the scattering operator into the formula for scattering cross section, one obtains three terms in the response function $\chi(q,\Omega)$: a nonresonant term; a mixed term; and a pure resonant term.  The result is
\begin{align}\label{eq13}
    R(\bm q,\Omega)=\frac{2\pi g^{2}(\bm k_{i})g^{2}(\bm k_{f})}{1-\exp(-\beta \Omega)}
    \chi(\bm q,\Omega),
\end{align}
where
\begin{align}\label{eq14}
    \chi(\bm q,\Omega)=\chi_{N}(\bm q,\Omega)+\chi_{M}(\bm q,\Omega)+\chi_{R}(\bm q,\Omega).
\end{align}
In Ref.~\onlinecite{sh_v_f_d2}, we have described in detail how to extract the components of the cross section from the appropriate correlation functions in the normal phase: we must calculate corresponding multi-time correlation functions for imaginary Matsubara frequencies and then analytically continue to the real axis. 
Inelastic light scattering examines charge excitations of different symmetries by employing polarizers on both the incident and scattered light. The $A_{\textrm{1g}}$ symmetry has the full symmetry of the lattice and is primarily measured by taking the initial and final polarizations to be ${\bm e}^{i}={\bm e}^{f}=(1,1,1,1,\ldots)$. The $B_{\textrm{1g}}$ symmetry involves crossed polarizers: ${\bm e}^{i}=(1,1,1,1,\ldots)$ and ${\bm e}^{f}=(-1,1,-1,1,\ldots)$; while the $B_{\textrm{2g}}$ symmetry is also using crossed polarizers, but with the polarizers rotated by 45 degrees; it requires the polarization vectors to satisfy ${\bm e}^{i}=(\sqrt{2},0,\sqrt{2},0,\ldots)$ and ${\bm e}^{f}=(0,\sqrt{2},0,\sqrt{2},\ldots)$. (Note in previous work we used the wrong normalization for the $B_{\rm 2g}$ polarization vectors resulting in a resonant response a factor of four smaller). For Raman scattering  ($\bf q=0$), it is easy to show that for a system with only nearest-neighbor hopping and in the limit of large spatial dimensions, the $A_{\textrm{1g}}$ sector has contributions from nonresonant, mixed and resonant scattering, the $B_{\textrm{1g}}$ sector has contributions from nonresonant and resonant scattering only, and the $B_{\textrm{2g}}$ sector is purely resonant~\cite{freericks_deveraux1,freericks_deveraux2}.  These results continue to hold in the ordered phase.

\section{Mixed and resonant contributions to the scattering response}

\begin{figure}[tb]
\noindent\includegraphics[scale=.8]{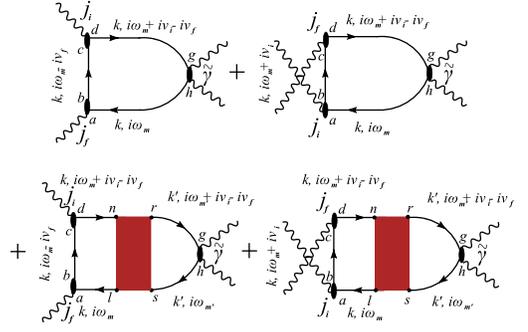}
   \caption{Feynman diagrams for the generalized polarizations of the mixed response function.
   Due to the static nature of the irreducible charge vertex of the Falicov-Kimball model, we have
   $i\omega_m=i\omega_{m^\prime}$.}\label{fig:diagr_mix}
\end{figure}

Since the nonresonant contributions to Raman scattering in the ordered phase have already been determined~\cite{MSF2}, we focus here on the modifications needed in the ordered phase to calculate the mixed and resonant responses.
As discussed above, the mixed and resonant response functions are extracted from the corresponding multi-time correlation functions. For the mixed one, the appropriate response function is built on the stress tensor and two current operators, as follows
\begin{equation}\label{eq15}
 \chi_{\tilde{\gamma},f,i}(\tau_{1},\tau_{2},\tau_{3})=\left\langle
   T_{\tau}\tilde{\gamma}(\tau_{1})j^{(f)}(\tau_{2})j^{(i)}(\tau_{3})\right\rangle.
\end{equation}
The symbol $T_{\tau}$ is a time ordering operator ($\langle\dots\rangle=\mathrm{Tr}[e^{-\beta\hat H}\dots]/\mathcal{Z}$). Here we have introduced a compact notation for the contraction of the stress tensor and current operators [Eqs.~(\ref{eq:current}) and (\ref{eq:stress}) for $\bm q=0$, $\bm k_{i(f)}=0$] with the polarization vectors, as follows:
\begin{align}\label{eq16}
&\tilde{\gamma}=\sum\limits_{\alpha\beta}e_{\alpha}^{i}\gamma_{\alpha,\beta}e_{\beta}^{f},\nonumber \\
&j^{(i)}=\sum\limits_{\alpha}e_{\alpha}^{i}j_{\alpha},\\
&j^{(f)}=\sum\limits_{\alpha}e_{\alpha}^{f}j_{\alpha} \nonumber,
\end{align}
respectively. The next step is to perform the Fourier transformation from imaginary time to imaginary Matsubara frequency, and, as a result, the mixed correlation function is represented as a sum over Matsubara frequencies of the generalized polarizations as follows:
\begin{align}\label{eq17}
  &\chi_{\tilde{\gamma},f,i}(i\nu_i-i\nu_f,i\nu_f,-i\nu_i) \\
  &=T\sum\limits_{m}\left [\Pi_{m-f,m+i-f,m}^{M}+\Pi_{m+i,m+i-f,m}^{M}\right
  ].\nonumber
\end{align}
Here we introduce the shorthand notation $\Pi_{m-f,m+i-f,m}^{M}=\Pi^{M}(i\omega_{m}-i\nu_{f},i\omega_{m}+i\nu_{i}-i\nu_{f},i\omega_{m})$ for the dependence on the fermionic $i\omega_{m}=i\pi T(2m+1)$ and bosonic $i\nu_{l}=i2\pi Tl$ Matsubara frequencies. The corresponding Feynman diagrams for the generalized contributions to the mixed response function are shown in Fig.~\ref{fig:diagr_mix}, where the first and third diagrams correspond to the first term in Eq.~(\ref{eq17}) and the other two diagrams correspond to the second one.

\begin{figure}[htb]
\noindent \includegraphics[scale=.8]{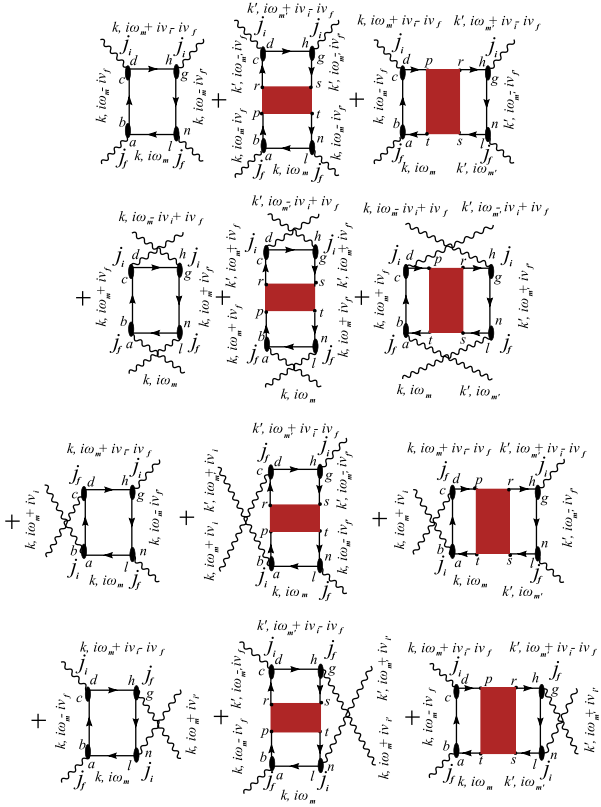}
   \caption{Feynman diagrams for the generalized polarizations of the resonant response function.}
   \label{fig:diagr_res}
\end{figure}

For the resonant response function, we construct the four-time correlation function with four current operators as follows
\begin{equation}\label{eq18}
 \chi_{i,f,f,i}(\tau_{1},\tau_{2},\tau_{3},\tau_{4})=\left\langle
   T_{\tau}j^{(i)}(\tau_{1})j^{(f)}(\tau_{2})j^{(f)}(\tau_{3})j^{(i)}(\tau_{4})\right\rangle,
\end{equation}
and in the same way as for the mixed one, the resonant response function is expressed as a sum of the generalized polarizations over Matsubara frequencies
\begin{align}\label{eq19}
  &\chi_{i,f,f,i}(-i\nu_i,i\nu_f,-i\nu_{f}',i\nu_{i}')=T\sum\limits_{m} \\
  &\times\bigl[\Pi_{m,m-f,m+i-f,m-f'}^{R,I}+\Pi_{m,m+f',m-i+f,m+f}^{R,I} \nonumber \\
  &+\Pi_{m,m+i,m+i-f,m-f'}^{R,II}+\Pi_{m,m-f,m+i-f,m+i'}^{R,II}\bigr].
  \nonumber
\end{align}
The corresponding Feynman diagrams for the generalized contributions to the resonant response function are shown in Fig.~\ref{fig:diagr_res}, where we introduce additional sublattice indices $a$ to $s$. Each term in Eq.~(\ref{eq19}) corresponds to a separate line in Fig.~\ref{fig:diagr_res}, respectively. There are also other contributions to the four-time correlation function in Eq.~(\ref{eq18}), but they do not contribute to the scattering cross section (see Ref.~\onlinecite{sh_v_f_d2} for details). For the $B_{\textrm{1g}}$ and $B_{\textrm{2g}}$ symmetries, the generalized polarization $\Pi_{m,m-f,m+i-f,m-f'}^{R,I}$ is a sum of the first two diagrams in the first line of Fig.~\ref{fig:diagr_res} (the bare loop and the vertical renormalization) and the generalized polarization $\Pi_{m,m+i,m+i-f,m-f'}^{R,II}$ contains only the first diagram in the third line (the bare loop), whereas for the $A_{\textrm{1g}}$ symmetry, all diagrams in the corresponding lines (the bare loop, the vertical renormalization, and the horizontal renormalization) contribute.

The next step is to derive analytic expressions for these mixed and resonant generalized polarizations. There are two types of Feynman diagrams (for both the mixed and resonant contributions as well as for the nonresonant one~\cite{MSF2}): bare loops and renormalized loops (see Figs.~\ref{fig:diagr_mix} and \ref{fig:diagr_res}). First we consider the bare loops and then the renormalized ones. The bare term for the mixed response $\Pi_{1,2,3}^{M,b}$ in the CDW phase is equal to
  \begin{align}\label{eq20}
  &\Pi_{1,2,3}^{M,b}
  =\frac{1}{N}\sum_{\bm k} j_{\bm k}^{(i)}j_{\bm k}^{(f)}\bar{\gamma}_{\bm k}
  \\
  \times &\biggl(G_{\bm k, 1}^{AA}G_{\bm k, 2}^{BA}G_{\bm k, 3}^{BB}
  +G_{\bm k, 1}^{AA}G_{\bm k, 2}^{BB}G_{\bm k, 3}^{AB}
  +G_{\bm k, 1}^{AB}G_{\bm k, 2}^{AA}G_{\bm k, 3}^{BB}
  \nonumber\\
  &+G_{\bm k, 1}^{AB}G_{\bm k, 2}^{AB}G_{\bm k, 3}^{AB}
  +G_{\bm k, 1}^{BA}G_{\bm k, 2}^{BA}G_{\bm k, 3}^{BA}
  +G_{\bm k, 1}^{BA}G_{\bm k, 2}^{BB}G_{\bm k, 3}^{AA}
  \nonumber\\
  &+G_{\bm k, 1}^{BB}G_{\bm k, 2}^{AA}G_{\bm k, 3}^{BA}
  +G_{\bm k, 1}^{BB}G_{\bm k, 2}^{AB}G_{\bm k, 3}^{AA}\biggr)
  \nonumber \\
  =&\dfrac{1}{2}\biggl[(i\omega_1+\mu_{d}^{B}-\Sigma_{1}^B)
  (i\omega_3+\mu_{d}^{A}-\Sigma_{3}^A)+(i\omega_1+\mu_{d}^{B}-\Sigma_{1}^B)
  \nonumber\\
  &\times(i\omega_2+\mu_{d}^{A}-\Sigma_{2}^A)
  +(i\omega_2+\mu_{d}^{B}-\Sigma_{2}^B)(i\omega_3+\mu_{d}^{A}
  -\Sigma_{3}^A)
    \nonumber\\
  &+(i\omega_2+\mu_{d}^{A}-\Sigma_{2}^A)(i\omega_3+\mu_{d}^{B}
  -\Sigma_{3}^B)+(i\omega_1+\mu_{d}^{A}-\Sigma_{1}^A)
    \nonumber\\
  &\times(i\omega_2+\mu_{d}^{B}-\Sigma_{2}^B)
  +(i\omega_1+\mu_{d}^{A}-\Sigma_{1}^A)(i\omega_3+\mu_{d}^{B}
  -\Sigma_{3}^B)\biggr]
  \nonumber \\
  &\times\biggl[\dfrac{\bar{Z}_{1}F_{\infty}(\bar{Z}_{1})}
  {(\bar{Z}^{2}_{2}-\bar{Z}^{2}_{1})(\bar{Z}^{2}_{3}-\bar{Z}^{2}_{1})}
  +\dfrac{\bar{Z}_{2}F_{\infty}(\bar{Z}_{2})}
  {(\bar{Z}^{2}_{1}-\bar{Z}^{2}_{2})(\bar{Z}^{2}_{3}-\bar{Z}^{2}_{2})}
    \nonumber \\
  &+\dfrac{\bar{Z}_{3}F_{\infty}(\bar{Z}_{3})}
  {(\bar{Z}^{2}_{1}-\bar{Z}^{2}_{3})(\bar{Z}^{2}_{2}-\bar{Z}^{2}_{3})}\biggr]
    \nonumber\\
  &+\biggl(\dfrac{\bar{Z}_{1}^{3}F_{\infty}(\bar{Z}_{1})}
  {(\bar{Z}^{2}_{2}-\bar{Z}^{2}_{1})(\bar{Z}^{2}_{3}-\bar{Z}^{2}_{1})}
  +\dfrac{\bar{Z}_{2}^{3}F_{\infty}(\bar{Z}_{2})}
  {(\bar{Z}^{2}_{1}-\bar{Z}^{2}_{2})(\bar{Z}^{2}_{3}-\bar{Z}^{2}_{2})}
    \nonumber \\
  &+\dfrac{\bar{Z}_{3}^{3}F_{\infty}(\bar{Z}_{3})}
  {(\bar{Z}^{2}_{1}-\bar{Z}^{2}_{3})(\bar{Z}^{2}_{2}-\bar{Z}^{2}_{3})}\biggr).
  \nonumber
  \end{align}
Here, we use the shorthand notation for frequencies: $i\omega_1$, $i\omega_2$, and $i\omega_3\to 1$, 2, and 3, with $j_{\bm k}^{(i(f))}=\sum_{\alpha}e_{\alpha}^{i(f)} \frac{\partial \epsilon_{\bm k}}{\partial k_{\alpha}}$ and $\bar{\gamma}_{\bm k}=\sum_{\alpha \beta}e_{\alpha}^{i} \frac{\partial^{2} \epsilon_{\bm k}}{\partial k_{\alpha} \partial k_{\beta}}e_{\beta}^{f}$, and $\bar Z(\omega)$ defined by
\begin{equation}\label{eq23}
  \bar{Z}(\omega)=\sqrt{[\omega+\mu^{A}_{d}-\Sigma^{A}(\omega)][\omega+\mu^{B}_{d}-\Sigma^{B}(\omega)]},
\end{equation}
where
\begin{equation}\label{eq24}
  F_\infty[\bar{Z}(\omega)]=\int d\epsilon
  \rho(\epsilon)\frac{1}{\bar{Z}(\omega)-\epsilon}
\end{equation}
is the Hilbert transform of the noninteracting density of states, which satisfies $\rho(\epsilon)=\exp(-\epsilon^{2}/t^{*2})/t^*\sqrt{\pi}$ for the infinite-dimensional hypercubic lattice.

The bare loop for the resonant response $\Pi_{1,2,3,4}^{R, b}$ (Fig.~\ref{fig:diagr_res}) in the CDW phase is equal to
  \begin{align}\label{eq25}
  &\Pi_{1,2,3,4}^{R,b}
  =\frac{1}{N}\sum_{\bm k}
  \sum_{\genfrac{}{}{0pt}{}{a\ne b}{c\ne d} }^{A,B}
  \sum_{\genfrac{}{}{0pt}{}{h\ne g}{l\ne n} }^{A,B}
  j_{\bm k}^{(i)}j_{\bm k}^{(f)}j_{\bm k}^{(i)}j_{\bm k}^{(f)}
  \nonumber \\
  \times &G_{\bm k, 1}^{na}G_{\bm k, 2}^{bc}G_{\bm k, 3}^{dh}G_{\bm k, 3}^{gl}
  \nonumber \\
  =&\dfrac{1}{4}\biggl\{\Bigl([i\omega_1+\mu_{d}^{A}-\Sigma_{1}^A][i\omega_2+\mu_{d}^{B}-\Sigma_{2}^B]
  \\
  &\times[i\omega_3+\mu_{d}^{A}-\Sigma_{3}^A][i\omega_4+\mu_{d}^{B}-\Sigma_{4}^B]
  \nonumber \\
  &+[i\omega_1+\mu_{d}^{B}-\Sigma_{1}^B][i\omega_2+\mu_{d}^{A}-\Sigma_{2}^A]
  \nonumber \\
  &\times[i\omega_3+\mu_{d}^{B}-\Sigma_{3}^B][i\omega_4+\mu_{d}^{A}-\Sigma_{4}^A]\Bigr)
  \chi_{1}(\bar{Z}_{1},\bar{Z}_{2},\bar{Z}_{3},\bar{Z}_{4})
    \nonumber\\
  +&\sum_{\nu \upsilon}^{\omega_1\dots\omega_4}
  [i\nu+\mu_{d}^{A}-\Sigma_{\nu}^A][i\upsilon+\mu_{d}^{B}-\Sigma_{\upsilon}^B]
  \chi'_{1}(\bar{Z}_{1},\bar{Z}_{2},\bar{Z}_{3},\bar{Z}_{4})
    \nonumber\\
  +&2\chi''_{1}(\bar{Z}_{1},\bar{Z}_{2},\bar{Z}_{3},\bar{Z}_{4})\biggr\}.
  \nonumber
  \end{align}
Here we introduce three quantities $\chi_{1}(\bar{Z}_{1},\bar{Z}_{2},\bar{Z}_{3},\bar{Z}_{4})$, $\chi'_{1}(\bar{Z}_{1},\bar{Z}_{2},\bar{Z}_{3},\bar{Z}_{4})$, and $\chi''_{1}(\bar{Z}_{1},\bar{Z}_{2},\bar{Z}_{3},\bar{Z}_{4})$, which are equal to
  \begin{align}\label{eq:chi1}
  &\chi_{1}(\bar{Z}_{1},\bar{Z}_{2},\bar{Z}_{3},\bar{Z}_{4})
   \\
  =&\dfrac{1}{N}\sum_{\bm k}\dfrac{1}{(\bar{Z}_{1}^{2}-\epsilon_{\bm k}^{2})
  (\bar{Z}_{2}^{2}-\epsilon_{\bm k}^{2})(\bar{Z}_{3}^{2}-\epsilon_{\bm k}^{2})
  (\bar{Z}_{4}^{2}-\epsilon_{\bm k}^{2})}
    \nonumber \\
  =&\dfrac{F_{\infty}(\bar{Z}_{1})/\bar{Z}_{1}}
  {(\bar{Z}^{2}_{2}-\bar{Z}^{2}_{1})(\bar{Z}^{2}_{3}-\bar{Z}^{2}_{1})(\bar{Z}^{2}_{4}-\bar{Z}^{2}_{1})}
    \nonumber \\
  &+\dfrac{F_{\infty}(\bar{Z}_{2})/\bar{Z}_{2}}
  {(\bar{Z}^{2}_{1}-\bar{Z}^{2}_{2})(\bar{Z}^{2}_{3}-\bar{Z}^{2}_{2})(\bar{Z}^{2}_{4}-\bar{Z}^{2}_{2})}
    \nonumber \\
  &+\dfrac{F_{\infty}(\bar{Z}_{3})/\bar{Z}_{3}}
  {(\bar{Z}^{2}_{1}-\bar{Z}^{2}_{3})(\bar{Z}^{2}_{2}-\bar{Z}^{2}_{3})(\bar{Z}^{2}_{4}-\bar{Z}^{2}_{3})}
    \nonumber \\
  &+\dfrac{F_{\infty}(\bar{Z}_{4})/\bar{Z}_{4}}
  {(\bar{Z}^{2}_{1}-\bar{Z}^{2}_{4})(\bar{Z}^{2}_{2}-\bar{Z}^{2}_{4})(\bar{Z}^{2}_{3}-\bar{Z}^{2}_{4})},
  \nonumber
  \end{align}
  \begin{align}\label{eq:chi1p}
  &\chi'_{1}(\bar{Z}_{1},\bar{Z}_{2},\bar{Z}_{3},\bar{Z}_{4})
    \\
  =&\dfrac{1}{N}\sum_{\bm k}\dfrac{\epsilon_{\bm k}^{2}}{(\bar{Z}_{1}^{2}-\epsilon_{\bm k}^{2})
  (\bar{Z}_{2}^{2}-\epsilon_{\bm k}^{2})(\bar{Z}_{3}^{2}-\epsilon_{\bm k}^{2})
  (\bar{Z}_{4}^{2}-\epsilon_{\bm k}^{2})}
    \nonumber \\
  =&\dfrac{\bar{Z}_{1}F_{\infty}(\bar{Z}_{1})}
  {(\bar{Z}^{2}_{2}-\bar{Z}^{2}_{1})(\bar{Z}^{2}_{3}-\bar{Z}^{2}_{1})(\bar{Z}^{2}_{4}-\bar{Z}^{2}_{1})}
    \nonumber \\
  &+\dfrac{\bar{Z}_{2}F_{\infty}(\bar{Z}_{2})}
  {(\bar{Z}^{2}_{1}-\bar{Z}^{2}_{2})(\bar{Z}^{2}_{3}-\bar{Z}^{2}_{2})(\bar{Z}^{2}_{4}-\bar{Z}^{2}_{2})}
    \nonumber \\
  &+\dfrac{\bar{Z}_{3}F_{\infty}(\bar{Z}_{3})}
  {(\bar{Z}^{2}_{1}-\bar{Z}^{2}_{3})(\bar{Z}^{2}_{2}-\bar{Z}^{2}_{3})(\bar{Z}^{2}_{4}-\bar{Z}^{2}_{3})}
    \nonumber \\
  &+\dfrac{\bar{Z}_{4}F_{\infty}(\bar{Z}_{4})}
  {(\bar{Z}^{2}_{1}-\bar{Z}^{2}_{4})(\bar{Z}^{2}_{2}-\bar{Z}^{2}_{4})(\bar{Z}^{2}_{3}-\bar{Z}^{2}_{4})},
  \nonumber
  \end{align}
  and
  \begin{align}\label{eq:chi1pp}
  &\chi''_{1}(\bar{Z}_{1},\bar{Z}_{2},\bar{Z}_{3},\bar{Z}_{4})
    \\
  =&\dfrac{1}{N}\sum_{\bm k}\dfrac{\epsilon_{\bm k}^{4}}{(\bar{Z}_{1}^{2}-\epsilon_{\bm k}^{2})
  (\bar{Z}_{2}^{2}-\epsilon_{\bm k}^{2})(\bar{Z}_{3}^{2}-\epsilon_{\bm k}^{2})
  (\bar{Z}_{4}^{2}-\epsilon_{\bm k}^{2})}
    \nonumber \\
  =&\dfrac{\bar{Z}_{1}^{3}F_{\infty}(\bar{Z}_{1})}
  {(\bar{Z}^{2}_{2}-\bar{Z}^{2}_{1})(\bar{Z}^{2}_{3}-\bar{Z}^{2}_{1})(\bar{Z}^{2}_{4}-\bar{Z}^{2}_{1})}
    \nonumber \\
  &+\dfrac{\bar{Z}_{2}^{3}F_{\infty}(\bar{Z}_{2})}
  {(\bar{Z}^{2}_{1}-\bar{Z}^{2}_{2})(\bar{Z}^{2}_{3}-\bar{Z}^{2}_{2})(\bar{Z}^{2}_{4}-\bar{Z}^{2}_{2})}
    \nonumber \\
  &+\dfrac{\bar{Z}_{3}^{3}F_{\infty}(\bar{Z}_{3})}
  {(\bar{Z}^{2}_{1}-\bar{Z}^{2}_{3})(\bar{Z}^{2}_{2}-\bar{Z}^{2}_{3})(\bar{Z}^{2}_{4}-\bar{Z}^{2}_{3})}
    \nonumber \\
  &+\dfrac{\bar{Z}_{4}^{3}F_{\infty}(\bar{Z}_{4})}
  {(\bar{Z}^{2}_{1}-\bar{Z}^{2}_{4})(\bar{Z}^{2}_{2}-\bar{Z}^{2}_{4})(\bar{Z}^{2}_{3}-\bar{Z}^{2}_{4})},
  \nonumber
  \end{align}
respectively.

The renormalized loops in the Feynman diagrams in Figs.~\ref{fig:diagr_mix} and \ref{fig:diagr_res} describe the charge screening effects through the reducible charge vertex, which is defined through the irreducible one by a Bethe-Salpeter equation. In the DMFT approach, the irreducible charge vertex ${\Gamma}_{a}$ is local but is different for different sublattices in the CDW ordered phase (see Ref.~\onlinecite{MSF2}). Nevertheless, it has the same functional form (when expressed as a functional of the Green's function and self-energy) as in the normal state~\cite{charge_vertex1,charge_vertex2,charge_vertex3} and is equal to
\begin{align}\label{eq: Gamma}
  &\Gamma_{a}(i\omega_{m},i\omega_{m'};i\nu_{l})
  =\delta_{m m'}\Gamma^{a}_{m,m+l}
  \\
  &\Gamma^{a}_{m,m+l}=\dfrac{1}{T}\dfrac{\Sigma^{a}_{m}-\Sigma^{a}_{m+l}}
   {G^{aa}_{m}-G^{aa}_{m+l}}
  \nonumber
\end{align}
for the Falicov-Kimball model (an explicit formula for other models is unknown). This expression also follows from the partially integrated Ward identity derived by Janis~\cite{janis}. Because in the CDW  phase the irreducible charge vertex is local both in the lattice and sublattice indices, the reducible one depends on two sublattice indices and is defined by the Bethe-Salpeter equation
\begin{equation}\label{eq28}
  \tilde{\Gamma}^{ab}_{m,m+l}=\delta_{ab}\Gamma^{a}_{m,m+l}
  +T\Gamma^{a}_{m,m+l} \sum_c \chi^{ac}_{m,m+l}
  \tilde{\Gamma}^{cb}_{m,m+l},
\end{equation}
where we introduce the bare susceptibility
\begin{equation}\label{eq29}
  \chi^{ab}_{m,m+l}= -\frac{1}{N}\sum_{\bm k} G^{ab}_{\bm k, m} G^{ba}_{\bm k, m+l}.
\end{equation}
The lattice Green functions can be derived from the Dyson equation in Eq.~(\ref{Dyson}) and are equal to
\begin{align}
 G_{\bm k,m}^{AA}&=\frac{i\omega_m+\mu-\Sigma_m^B}{\bar{Z}_{m}^{2}-\epsilon_{\bm k}^{2}},
 \\
 G_{\bm k,m}^{BB}&=\frac{i\omega_m+\mu-\Sigma_m^A}{\bar{Z}_{m}^{2}-\epsilon_{\bm k}^{2}},
 \nonumber\\
 G_{\bm k,m}^{AB}&=G_{\bm k,m}^{BA}=\frac{\epsilon_{\bm k}}{\bar{Z}_{m}^{2}-\epsilon_{\bm k}^{2}}.
 \nonumber 
\end{align}
Expressions for the renormalized loops have a similar form for all contributions (nonresonant, mixed, and resonant) and differ only in the loops attached to the left and right sides of the total reducible charge vertex. The renormalized loop for the mixed response is then equal to
\begin{align}\label{eq30}
   \Pi_{1,2,3}^{M,r}&=
  \left[
   \begin{array}{cc}
   \chi_{jj}^{A}(i\omega_1,i\omega_2,i\omega_3)
   & \chi_{jj}^{B}(i\omega_1,i\omega_2,i\omega_3) \\
   \end{array}
  \right] \\
  &\times T\left\|
   \begin{array}{cc}
   \tilde{\Gamma}^{AA}_{1,3} & \tilde{\Gamma}^{AB}_{1,3} \\
   \tilde{\Gamma}^{BA}_{1,3} & \tilde{\Gamma}^{BB}_{1,3} \\
   \end{array}
  \right\|
\left[
   \begin{array}{c} \chi_{\bar{\gamma}}^{A}(i\omega_1,i\omega_3) \\
   \chi_{\bar{\gamma}}^{B}(i\omega_1,i\omega_3)
   \end{array}
  \right],
  \nonumber
\end{align}
where we introduce the quantities
\begin{align}\label{eq31}
  &\chi_{jj}^{A}(i\omega_1,i\omega_2,i\omega_3)=\frac1N\sum_{\bm k}j_{\bm k}^{(i)}j_{\bm k}^{(f)}
  \\
  &\times\bigl[G_{\bm k,\omega_1}^{AA}G_{\bm k,\omega_2}^{BA}G_{\bm k,\omega_3}^{BA}
  +G_{\bm k,\omega_1}^{AA}G_{\bm k,\omega_2}^{BB}G_{\bm k,\omega_3}^{AA}
  \nonumber \\
  &+G_{\bm k,\omega_1}^{AB}G_{\bm k,\omega_2}^{AA}G_{\bm k,\omega_3}^{BA}
  +G_{\bm k,\omega_1}^{AB}G_{\bm k,\omega_2}^{AB}G_{\bm k,\omega_3}^{AA}\bigr]
  \nonumber \\
  =&\bigl[i(\omega_1+\omega_2+\omega_3)+3\mu_d^B-\Sigma_1^B-\Sigma_2^B-\Sigma_3^B\bigr]
  \nonumber \\
  \times&\biggl(\dfrac{\bar{Z}_{1}F_{\infty}(\bar{Z}_{1})}
  {(\bar{Z}_{2}^{2}-\bar{Z}_{1}^{2})(\bar{Z}_{3}^{2}-\bar{Z}_{1}^{2})}
  +\dfrac{\bar{Z}_{2}F_{\infty}(\bar{Z}_{2})}
  {(\bar{Z}_{1}^{2}-\bar{Z}_{2}^{2})(\bar{Z}_{3}^{2}-\bar{Z}_{2}^{2})}
  \nonumber \\
  &+\dfrac{\bar{Z}_{1}F_{\infty}(\bar{Z}_{3})}
  {(\bar{Z}_{1}^{2}-\bar{Z}_{3}^{2})(\bar{Z}_{2}^{2}-\bar{Z}_{3}^{2})}\biggr)
  \nonumber \\
  +&\bigl[i\omega_1+\mu_d^B-\Sigma_1^B\bigr]\bigl[i\omega_2+\mu_d^A-\Sigma_2^A\bigr]\bigl[i\omega_3+\mu_d^B-\Sigma_3^B\bigr]
  \nonumber \\
  \times&\biggl(\dfrac{F_{\infty}(\bar{Z}_{1})/\bar{Z}_{1}}
  {(\bar{Z}_{2}^{2}-\bar{Z}_{1}^{2})(\bar{Z}_{3}^{2}-\bar{Z}_{1}^{2})}
  +\dfrac{F_{\infty}(\bar{Z}_{2})/\bar{Z}_{2}}
  {(\bar{Z}_{1}^{2}-\bar{Z}_{2}^{2})(\bar{Z}_{3}^{2}-\bar{Z}_{2}^{2})}
  \nonumber \\
  &+\dfrac{F_{\infty}(\bar{Z}_{3})/\bar{Z}_{3}}
  {(\bar{Z}_{1}^{2}-\bar{Z}_{3}^{2})(\bar{Z}_{2}^{2}-\bar{Z}_{3}^{2})}\biggr)
  \nonumber
\end{align}
and
\begin{align}\label{eq32}
  &\chi_{jj}^{B}(i\omega_1,i\omega_2,i\omega_3)=\frac1N\sum_{\bm k}j_{\bm k}^{(i)}j_{\bm k}^{(f)}
  \\
  &\times\bigl[G_{\bm k,\omega_1}^{BA}G_{\bm k,\omega_2}^{BA}G_{\bm k,\omega_3}^{BB}
  +G_{\bm k,\omega_1}^{BA}G_{\bm k,\omega_2}^{BB}G_{\bm k,\omega_3}^{AB}
  \nonumber \\
  &+G_{\bm k,\omega_1}^{BB}G_{\bm k,\omega_2}^{AA}G_{\bm k,\omega_3}^{BB}
  +G_{\bm k,\omega_1}^{BB}G_{\bm k,\omega_2}^{AB}G_{\bm k,\omega_3}^{AB}\bigr]
  \nonumber \\
  =&\bigl[i(\omega_1+\omega_2+\omega_3)+3\mu_d^A-\Sigma_1^A-\Sigma_2^A-\Sigma_3^A\bigr]
  \nonumber \\
  \times&\biggl(\dfrac{\bar{Z}_{1}F_{\infty}(\bar{Z}_{1})}
  {(\bar{Z}_{2}^{2}-\bar{Z}_{1}^{2})(\bar{Z}_{3}^{2}-\bar{Z}_{1}^{2})}
  +\dfrac{\bar{Z}_{2}F_{\infty}(\bar{Z}_{2})}
  {(\bar{Z}_{1}^{2}-\bar{Z}_{2}^{2})(\bar{Z}_{3}^{2}-\bar{Z}_{2}^{2})}
  \nonumber \\
  &+\dfrac{\bar{Z}_{1}F_{\infty}(\bar{Z}_{3})}
  {(\bar{Z}_{1}^{2}-\bar{Z}_{3}^{2})(\bar{Z}_{2}^{2}-\bar{Z}_{3}^{2})}\biggr)
  \nonumber \\
  +&\bigl[i\omega_1+\mu_d^A-\Sigma_1^A\bigr]\bigl[i\omega_2+\mu_d^B-\Sigma_2^B\bigr]\bigl[i\omega_3+\mu_d^A-\Sigma_3^A\bigr]
  \nonumber \\
  \times&\biggl(\dfrac{F_{\infty}(\bar{Z}_{1})/\bar{Z}_{1}}
  {(\bar{Z}_{2}^{2}-\bar{Z}_{1}^{2})(\bar{Z}_{3}^{2}-\bar{Z}_{1}^{2})}
  +\dfrac{F_{\infty}(\bar{Z}_{2})/\bar{Z}_{2}}
  {(\bar{Z}_{1}^{2}-\bar{Z}_{2}^{2})(\bar{Z}_{3}^{2}-\bar{Z}_{2}^{2})}
  \nonumber \\
  &+\dfrac{F_{\infty}(\bar{Z}_{3})/\bar{Z}_{3}}
  {(\bar{Z}_{1}^{2}-\bar{Z}_{3}^{2})(\bar{Z}_{2}^{2}-\bar{Z}_{3}^{2})}\biggr),
  \nonumber
\end{align}
to the left of the charge vertex with
\begin{align}\label{eq33}
  \chi_{\tilde{\gamma}}^{A}(i\omega_1,i\omega_3)&=\frac1N\sum_{\bm k}\bar{\gamma}_{\bm k}
  \left[G_{\bm k,\omega_1}^{AA}G_{\bm k,\omega_3}^{BA}
  +G_{\bm k,\omega_1}^{AB}G_{\bm k,\omega_3}^{AA}\right]
  \nonumber \\
  =&\bigl[i(\omega_1+\omega_3)+2\mu_d^B-\Sigma_1^B-\Sigma_3^B\bigr]
  \\
  &\times\left(\dfrac{\bar{Z}_{1}F_{\infty}(\bar{Z}_{1})-\bar{Z}_{3}F_{\infty}(\bar{Z}_{3})}
  {\bar{Z}_{3}^{2}-\bar{Z}_{1}^{2}}\right)
  \nonumber
\end{align}
and
\begin{align}\label{eq34}
  \chi_{\tilde{\gamma}}^{B}(i\omega_1,i\omega_3)&=\frac1N\sum_{\bm k}\bar{\gamma}_{\bm k}
  \left[G_{\bm k,\omega_1}^{BA}G_{\bm k,\omega_3}^{BB}
  +G_{\bm k,\omega_1}^{BB}G_{\bm k,\omega_3}^{AB}\right]
  \nonumber \\
  =&\bigl[i(\omega_1+i\omega_3)+2\mu_d^A-\Sigma_1^A-\Sigma_3^A\bigr]
  \\
  &\times\left(\dfrac{\bar{Z}_{1}F_{\infty}(\bar{Z}_{1})-\bar{Z}_{3}F_{\infty}(\bar{Z}_{3})}
  {\bar{Z}_{3}^{2}-\bar{Z}_{1}^{2}}\right)
  \nonumber
\end{align}
to the right of the charge vertex. Now we can find the exact expression for the vertex corrections defined by Eq.~(\ref{eq30}) with the following form
\begin{align}\label{eq35}
  &\Pi_{1,2,3}^{M,r}=\frac{1}{\Delta_{1,3}}
  \\
  &\times\Bigl[
  \chi_{jj}^{A}(i\omega_1,i\omega_2,i\omega_3) T\Gamma^A_{1,3}\chi^{AB}_{1,3}T\Gamma^B_{1,3}\chi_{\tilde{\gamma}}^{B}(i\omega_1,i\omega_3)
  \nonumber \\
  &+\chi_{jj}^{A}(i\omega_1,i\omega_2,i\omega_3)\left(1-T\Gamma^B_{1,3}\chi^{BB}_{1,3}\right) T\Gamma^A_{1,3}\chi_{\tilde{\gamma}}^{A}(i\omega_1,i\omega_3)
  \nonumber \\
  &+\chi_{jj}^{B}(i\omega_1,i\omega_2,i\omega_3)\left(1-T\Gamma^A_{1,3}\chi^{AA}_{1,3}\right) T\Gamma^B_{1,3}\chi_{\tilde{\gamma}}^{B}(i\omega_1,i\omega_3)
  \nonumber \\
  &+\chi_{jj}^{B}(i\omega_1,i\omega_2,i\omega_3) T\Gamma^B_{1,3}\chi^{BA}_{1,3}T\Gamma^A_{1,3}\chi_{\tilde{\gamma}}^{A}(i\omega_1,i\omega_3)
  \Bigr],
  \nonumber
\end{align}
where
\begin{align}\label{eq36}
  \Delta_{1,3}&=\left(1-T\Gamma^A_{1,3}\chi^{AA}_{1,3}\right)
   \left(1-T\Gamma^B_{1,3}\chi^{BB}_{1,3}\right)
  \\
  &-T\Gamma^A_{1,3}\chi^{AB}_{1,3}T\Gamma^B_{1,3}\chi^{BA}_{1,3}.
  \nonumber
\end{align}

For the resonant response function, the renormalized loops in Feynman diagrams are defined in the same way as the mixed one and in a compact form we have
\begin{align}\label{eq37}
  &\Pi_{1,2,3,4}^{R,r}=\frac{1}{\Delta_{1,3}}
  \\
  &\times\Bigl[
  \chi_{jj}^{A}(i\omega_1,i\omega_2,i\omega_3)T\Gamma^A_{1,3}\chi^{AB}_{1,3}T\Gamma^B_{1,3}\chi_{jj}^{B}(i\omega_3,i\omega_4,i\omega_1)
  \nonumber\\
  &+\chi_{jj}^{A}(i\omega_1,i\omega_2,i\omega_3)\left(1-T\Gamma^B_{1,3}\chi^{BB}_{1,3}\right) T\Gamma^A_{1,3}\chi_{jj}^{A}(i\omega_3,i\omega_4,i\omega_1)
  \nonumber\\
  &+\chi_{jj}^{B}(i\omega_1,i\omega_2,i\omega_3)\left(1-T\Gamma^A_{1,3}\chi^{AA}_{1,3}\right) T\Gamma^B_{1,3}\chi_{jj}^{B}(i\omega_3,i\omega_4,i\omega_1)
  \nonumber\\
  &+\chi_{jj}^{B}(i\omega_1,i\omega_2,i\omega_3)T\Gamma^B_{1,3}\chi^{BA}_{1,3}T\Gamma^A_{1,3}\chi_{jj}^{A}(i\omega_3,i\omega_4,i\omega_1)
  \Bigr].
  \nonumber
\end{align}
Now we have the same quantities $\chi_{jj}^{A}(i\omega_1,i\omega_2,i\omega_3)$ to the left and to the right of the charge vertex. For nonresonant scattering, the renormalized contributions have the same form with $\chi_{jj}^{a}(i\omega_1,i\omega_2,i\omega_3)$ replaced by $\chi_{\tilde{\gamma}}^{a}(i\omega_1,i\omega_3)$ (see Ref.~\onlinecite{MSF2}).

The total expression for the mixed generalized polarization is finally obtained as the sum of both the bare and renormalized contributions:
\begin{align}\label{eq38}
\Pi_{1,2,3}^{M}=\Pi_{1,2,3}^{M,b}+\Pi_{1,2,3}^{M,r}
\end{align}
on the imaginary axis.
Now we have to perform an analytic continuation to the real axis. First we replace the sum over Matsubara frequencies by an integral over the real axis. Next we analytically continue Matsubara frequencies to the real axis in the following order: first $i\nu_{i}-i\nu_{f}=i\nu'_{i}-i\nu'_{f}\to\Omega\pm i0^{+}$ followed by $i\nu_{i(f)}\to\omega_{i(f)}\pm i0^{+}$, $i\nu'_{i(f)}\to\omega'_{i(f)}\pm i0^{+}$, and finally $\Delta\omega=\omega'_{i}-\omega_{i}=\omega'_{f}-\omega_{f}\to0$ in Eq.~(\ref{eq17}). Then the mixed response function is expressed directly in terms of the generalized polarizations as
\begin{align}\label{eq:chiM}
  &\chi_{M}(\Omega)
  =\frac{1}{(2\pi i)^{2}}\int\limits_{-\infty}^{+\infty} d \omega
  \left[f(\omega)-f(\omega+\Omega)\right]
  \nonumber\\
  &\times\textrm{Re}\Bigl\{\Pi^{M}(\omega-\omega_{f}+i0^{+},\omega+\Omega+i0^{+},\omega-i0^{+})
  \nonumber\\
  &-\Pi^{M}(\omega-\omega_{f}+i0^{+},\omega+\Omega-i0^{+},\omega-i0^{+})
  \nonumber \\
  &+\Pi^{M}(\omega-\omega_{f}-i0^{+},\omega+\Omega+i0^{+},\omega-i0^{+})
  \nonumber\\
  &-\Pi^{M}(\omega-\omega_{f}-i0^{+},\omega+\Omega-i0^{+},\omega-i0^{+})
  \\
  &+\Pi^{M}(\omega+\omega_{i}+i0^{+},\omega+\Omega+i0^{+},\omega-i0^{+})
  \nonumber\\
  &-\Pi^{M}(\omega+\omega_{i}+i0^{+},\omega+\Omega-i0^{+},\omega-i0^{+})
  \nonumber \\
  &+\Pi^{M}(\omega+\omega_{i}-i0^{+},\omega+\Omega+i0^{+},\omega-i0^{+})
  \nonumber\\
  &-\Pi^{M}(\omega+\omega_{i}-i0^{+},\omega+\Omega-i0^{+},\omega-i0^{+})
  \Bigr\},\nonumber
\end{align}
where $f(\omega)=1\left/[\exp(\beta\omega)+1]\right.$ is the Fermi-Dirac
distribution function.  Since the imaginary-axis form of the response is expressed as a functional of the Green's functions and self-energies, one simply replaces the appropriate Matsubara frequency arguments by the real frequencies, according to the different terms listed above.  This is a tedious, but straightforward exercise to yield the final formulas, which are too cumbersome to include here.

For the resonant response function, an analytical continuation onto the real axis is more complicated, but the general approach remains the same and final expression is the following:
\begin{align}\label{eq:chiR}
  &\chi_{R}(q,\Omega)
  =\frac{1}{(2\pi i)^{2}}\int\limits_{-\infty}^{+\infty} d \omega
  \left[f(\omega)-f(\omega+\Omega)\right]
   \\
  &\times\Bigl\{\lim\limits_{\Delta\omega\to 0}\bigl[
  \nonumber \\  &
\Pi^{R,I}(\omega-i0^{+},\omega-\omega_{f}-i0^{+},
  \omega+\Omega+i0^{+},\omega-\omega_{f'}+i0^{+})
  \nonumber \\
  &-\Pi^{R,I}(\omega+i0^{+},\omega-\omega_{f}-i0^{+},\omega+\Omega+i0^{+},\omega-\omega_{f'}+i0^{+})
  \nonumber \\
  &+\Pi^{R,I}(\omega+i0^{+},\omega-\omega_{f}-i0^{+},\omega+\Omega-i0^{+},\omega-\omega_{f'}+i0^{+})
  \nonumber \\
  &-\Pi^{R,I}(\omega-i0^{+},\omega-\omega_{f}-i0^{+},\omega+\Omega-i0^{+},\omega-\omega_{f'}+i0^{+})
  \nonumber \\
  &+\Pi^{R,I}(\omega-i0^{+},\omega+\omega_{i'}-i0^{+},\omega+\Omega+i0^{+},\omega+\omega_{i}+i0^{+})
  \nonumber \\
  &-\Pi^{R,I}(\omega+i0^{+},\omega+\omega_{i'}-i0^{+},\omega+\Omega+i0^{+},\omega+\omega_{i}+i0^{+})
  \nonumber \\
  &+\Pi^{R,I}(\omega+i0^{+},\omega+\omega_{i'}-i0^{+},\omega+\Omega-i0^{+},\omega+\omega_{i}+i0^{+})
  \nonumber \\
  &-\Pi^{R,I}(\omega-i0^{+},\omega+\omega_{i'}-i0^{+},\omega+\Omega-i0^{+},\omega+\omega_{i}+i0^{+})
  \bigr]
  \nonumber \\
  &+2\textrm{Re}\bigl[\Pi^{R,II}(\omega-i0^{+},\omega+\omega_{i}+i0^{+},\omega+\Omega+i0^{+},\omega-\omega_{f}+i0^{+})
  \nonumber \\
  &-\Pi^{R,II}(\omega-i0^{+},\omega+\omega_{i'}+i0^{+},\omega+\Omega-i0^{+},\omega-\omega_{f}+i0^{+})
  \nonumber \\
  &+\Pi^{R,II}(\omega-i0^{+},\omega-\omega_{f}-i0^{+},\omega+\Omega+i0^{+},\omega+\omega_{i}-i0^{+})
  \nonumber \\
  &-\Pi^{R,II}(\omega-i0^{+},\omega-\omega_{f}-i0^{+},\omega+\Omega-i0^{+},\omega+\omega_{i}-i0^{+})
  \bigr]\Bigr\}.\nonumber
\end{align}

Now we can specify the different contributions to the resonant response in the different symmetry channels. In the $B_{\textrm{1g}}$ and $B_{\textrm{2g}}$ channels, the generalized polarizations $\Pi^{R,II}$ contain only the bare loop contributions (the first diagrams in the last two lines of Fig.~\ref{fig:diagr_res}):
\begin{align}\label{eq41}
&\Pi_{B_{\textrm{1g}},1,2,3,4}^{R,II}=\Pi_{1,2,3,4}^{R,b}\,,
\\
&\Pi_{B_{\textrm{2g}},1,2,3,4}^{R,II}=\Pi_{B_{\textrm{1g}},1,2,3,4}^{R,II}\,.
\nonumber
\end{align}
On the other hand, the generalized polarization $\Pi^{R,I}$ contains both the bare and vertically renormalized contributions (the first two diagrams in the first two lines of Fig.~\ref{fig:diagr_res}) in the $B_{\rm 1g}$ and $B_{\rm 2g}$ symmetry channels:
\begin{align}\label{eq42}
&\Pi_{B_{1g},1,2,3,4}^{R,I}=\Pi_{1,2,3,4}^{R,b}+\Pi_{1,2,3,4}^{R,r}\,,\\
&\Pi_{B_{2g},1,2,3,4}^{R,I}=\Pi_{B_{1g},1,2,3,4}^{R,I}\,.
\nonumber
\end{align}
In the $A_{1g}$ channel, all diagrams in Fig.~\ref{fig:diagr_res} contribute, hence
\begin{align}\label{eq43}
&\Pi_{A_{1g},1,2,3,4}^{R,I}=\Pi_{A_{1g},1,2,3,4}^{R,II}
\\
&=3\Pi_{1,2,3,4}^{R,b}+\Pi_{1,2,3,4}^{R,r}+\Pi_{2,3,4,1}^{R,r}.
\nonumber
\end{align}
It should be noted that some renormalized terms in Eq.~(\ref{eq:chiR}) contain nominal divergences in the limit $\Delta\omega\to0$ [connected with vanishing determinants in Eq.~(\ref{eq36}) which are found in the denominators of Eq.~(\ref{eq37})], but the contribution of these terms to the response is actually finite. In the case of the uniform phase of the Falicov-Kimball model, their contributions were calculated analytically using l'Hopital's rule~\cite{sh_v_f_d2}, but in the case of the CDW phase, the expressions are more cumbersome, so we calculate the limit $\Delta\omega\to0$ numerically.

\section{Numerical Results}

\begin{figure}[htb]
\noindent\includegraphics[scale=.35,clip]{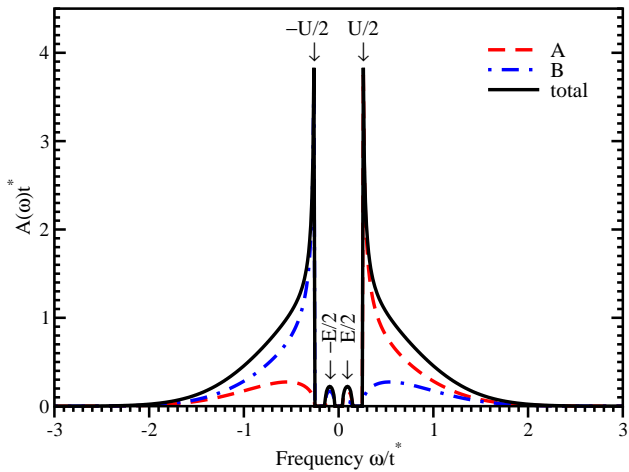}
   \caption{(Color online.) Conduction electron DOS at $T=0.02$ for $U=0.5$.  The solid black curve is the total DOS, while the dashed red line is for the $A$ sublattice and the dot-dashed blue line is for the $B$ sublattice. Note how there is a divergence at the band edge on each sublattice which develops as $T\rightarrow 0$, and that the subgap states disappear as $T\rightarrow 0$.  Finally, we have marked the locations of the band edge at $\pm U/2$, and of the peak of the subgap states at $\pm E/2$.}\label{fig:dos05}
\end{figure}

\begin{figure}[htb]
\noindent\includegraphics[scale=.35,clip]{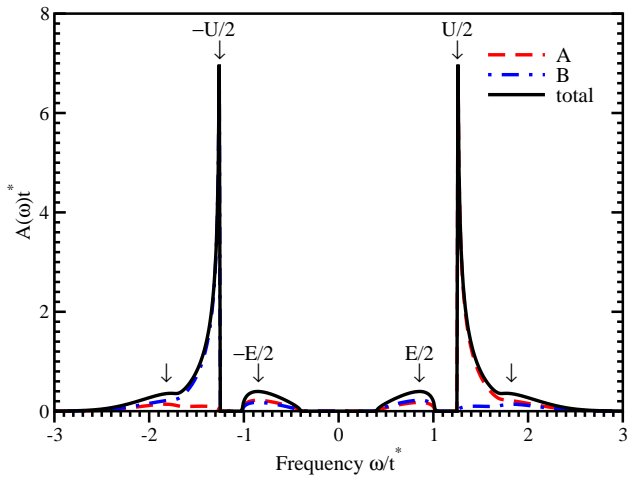}
   \caption{(Color online.) Conduction electron DOS at $T=0.06$ for $U=2.5$.  The solid black curve is the total DOS, while the dashed red line is for the $A$ sublattice and the dot-dashed blue line is for the $B$ sublattice.  Note how there is a divergence at the band edge on each sublattice which develops as $T\rightarrow 0$, and that the subgap states disappear as $T\rightarrow 0$.  Finally, we have marked the locations of the band edge at $\pm U/2$, and of the peak of the subgap states at $\pm E/2$. The DOS has upper and lower Mott shoulders for the strongly correlated system (indicated by unlabeled arrows).}\label{fig:dos25}
\end{figure}

Now that all of the formal developments are complete, we are ready to discuss the numerical results found by calculating the total Raman response function for different symmetry channels and different interaction strengths as functions of $T$ within the ordered phase.
We shall consider two cases: the case of a weakly scattering metal in the normal state ($U=0.5$, $T_{c}=0.0336$) and the case of a strongly correlated insulator in the normal state ($U=2.5$, $T_{c}=0.0724$); both cases are insulators at zero temperature due to the CDW order. In previous work~\cite{MSF1}, we have calculated the temperature evolution of the single particle DOS in the CDW phase of the Falicov-Kimball model. Here we present figures of the DOS for the temperatures that we calculate the Raman response (all temperatures are below $T_c$): $T=0.02$ for the case of $U=0.5$ (Fig.~\ref{fig:dos05}) and $T=0.06$ for the case of $U=2.5$ (Fig.~\ref{fig:dos25}), respectively.
One common feature of the CDW-ordered DOS is the presence of a sharp inverse square-root-like feature at $U/2$ for sublattice $A$ and $-U/2$ for sublattice $B$, which frame the gap as $T\rightarrow 0$. There also are bands of subgap states with a maximum DOS at $\pm E/2$; we have $E\approx0.18$ for $U=0.5$ and $E\approx1.7$ for $U=2.5$. These subgap states originate from thermal excitations of the CDW order and they vanish at zero temperature. In addition, for the case of the strongly correlated insulator $U=2.5$, one can observe in Fig.~\ref{fig:dos25} additional shoulders at $\pm1.82$, which are resulting from the upper and lower Hubbard bands of the high temperature normal state Mott insulator.  At zero temperature the states below zero energy are filled and the states above it are empty. At finite temperature, due to thermal occupation, there are some empty states below the chemical potential and some occupied ones above. These thermally activated states give contributions to the optical conductivity and to the nonresonant Raman scattering, creating different peaks in those functions~\cite{MSF1,MSF2}. A large main peak at $U$ corresponds to single-particle transitions from the lower occupied to the upper empty bands of the CDW, which are separated by a gap of width $U$. This peak will become more enhanced as $T\rightarrow 0$. Peaks also occur at $(U+E)/2$, which correspond to single-particle transitions from the lower occupied CDW band at $-U/2$ to the upper empty subgap states at $E/2$ and from the lower occupied subgap states at $-E/2$ to the upper empty CDW band at $U/2$. In addition, we see peaks at $E$ corresponding to transitions from the lower occupied subgap states at $-E/2$ to the upper empty subgap states at $E/2$. The intensity of these peaks will decrease as $T$ is lowered, since the subgap states will lose spectral weight, and eventually vanish. Finally, there is an additional peak at $(U-E)/2$, which corresponds to transitions between the almost fully occupied lower CDW band at $-U/2$ and the lower subgap states at $-E/2$ and between the almost empty upper subgap states at $E/2$ and the upper CDW band at $U/2$. The intensity of this peak will also shrink as $T$ is lowered.  We anticipate all of this structure will also to be seen in the total electronic Raman scattering, but the details of the temperature dependence, or of the resonant effects are difficult to guess without performing the calculations. We do see, however, that we have a wide number of different ``gap edges'' where one would expect large resonant effects.  The largest should occur when the photon energy is equal to $U$, but we should also see them at $(U\pm E)/2$ and $E$.

\begin{figure}[tb]
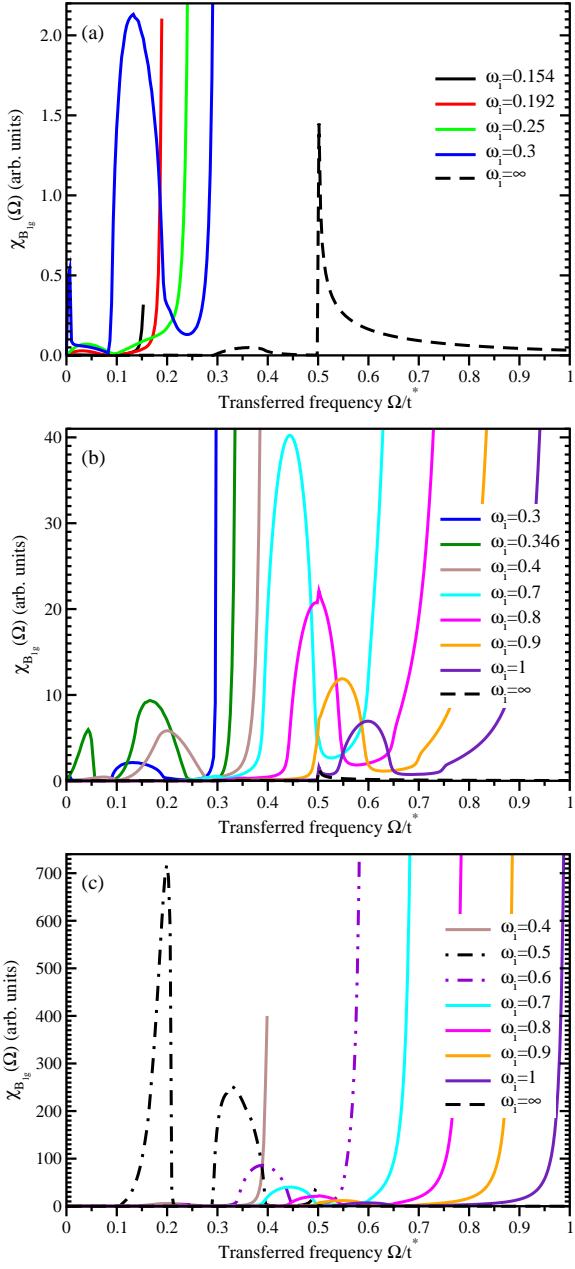

\noindent\includegraphics[scale=.31,clip]{fig5a}\\ [1ex]
\includegraphics[scale=.31,clip]{fig5b}\\ [1ex]
\includegraphics[scale=.31,clip]{fig5c}
   \caption{(Color online.) Total Raman spectra for $B_{\textrm{1g}}$ symmetry for different values of the incident photon frequency and for different vertical scales in the different panels with $T=0.02$ for $U=0.5$. Colors are used for the different incident frequencies, which can also be read off by examining the location of the unphysical divergence when $\Omega\rightarrow\omega_i$ on the hypercubic lattice. The nonresonant response (the case of $\omega_i=\infty$) is also shown with a dashed line.}\label{fig:spectra05_B1g}
\end{figure}

Analysis of the expression in Eq.~(\ref{eq:chiR}) gives that, in addition to the nonresonant peaks at $\Omega=U$, $(U+E)/2$, $E$, and $(U-E)/2$, there can also exist peaks which originate from two particle transitions, i.e. $\Omega=(3E-U)/2$, $(U+E)/2$, $E$, $U-E$, and $(U-E)/2$, some of which coincide with single particle transition energies. In addition, there can be strong resonant enhancement when either $\omega_i$ or $\omega_f$ approach these energies.

\begin{figure}[tb]
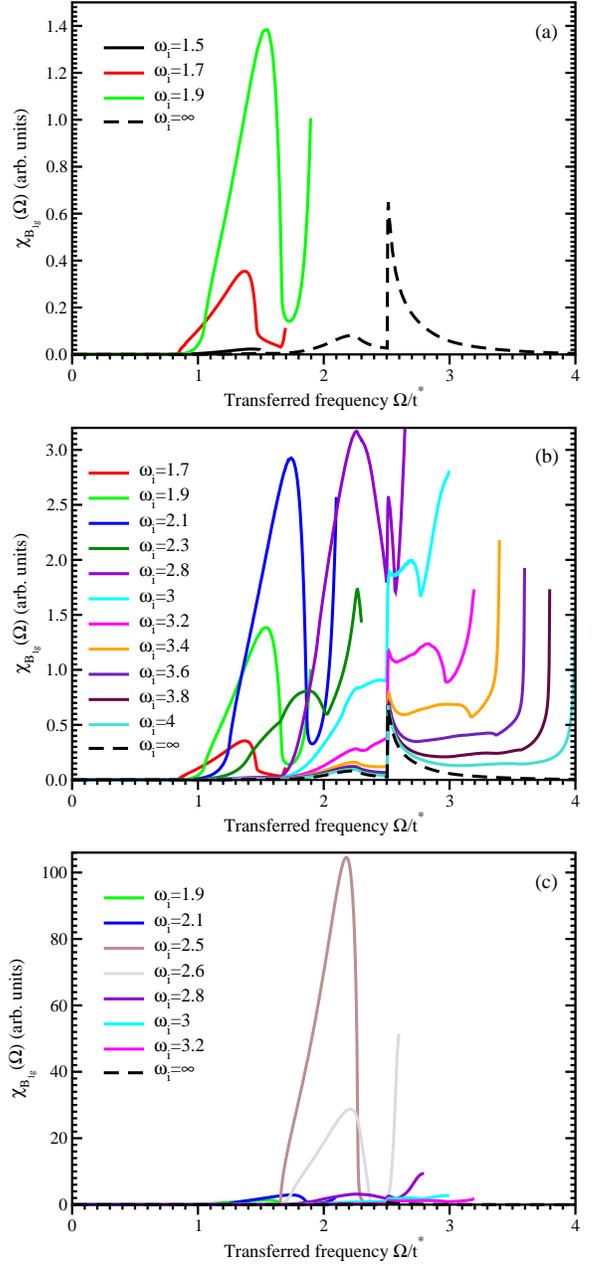

\noindent\includegraphics[scale=.31,clip]{fig6a}\\ [1ex]
\includegraphics[scale=.31,clip]{fig6b}\\ [1ex]
\includegraphics[scale=.31,clip]{fig6c}
   \caption{(Color online.) Total Raman spectra for $B_{\textrm{1g}}$ symmetry for different values of the incident photon frequency and for different vertical scales in the different panels with $T=0.06$ for $U=2.5$. Colors are used for the different incident frequencies, which can also be read off by examining the location of the unphysical divergence when $\Omega\rightarrow\omega_i$ on the hypercubic lattice. The nonresonant response (the case of $\omega_i=\infty$) is also shown with a dashed line.}\label{fig:spectra25_B1g}
\end{figure}

\begin{figure}[tb]
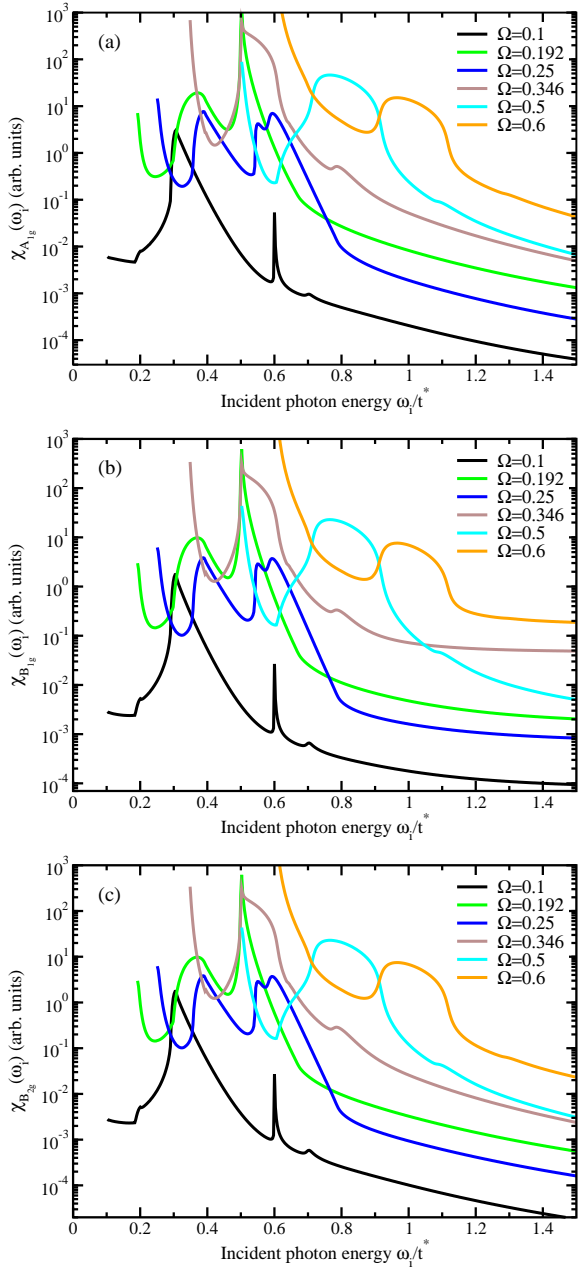

\noindent\includegraphics[scale=.31,clip]{fig7a}\\ [1ex]
\includegraphics[scale=.31,clip]{fig7b}\\ [1ex]
\includegraphics[scale=.31,clip]{fig7c}
   \caption{(Color online.) Resonant profiles in a semilog plot for different values of the transferred photon frequency at $T=0.02$ for $U=0.5$. Different colors denote different transferred frequency $\Omega$ (which can also be found from the unphysical divergence at $\omega_i\rightarrow \Omega$). Note how similar the response is for different symmetry channels.}\label{fig:profile05}
\end{figure}

\begin{figure}[tb]
\noindent\includegraphics[scale=.33,clip]{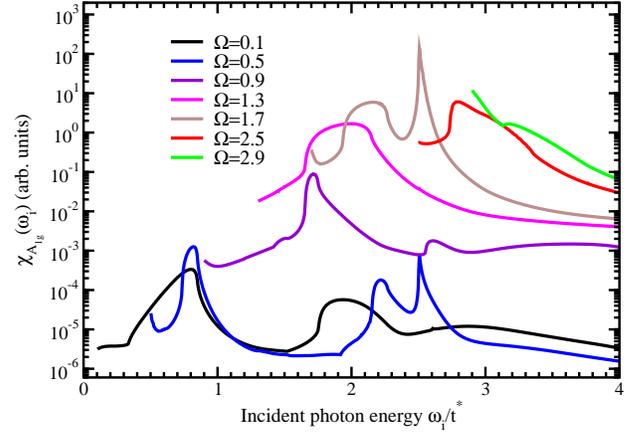}
   \caption{(Color online.) Resonant profiles in a semilog plot for different values of the transferred photon frequency at $T=0.06$ for $U=2.5$. Different colors denote different transferred frequency $\Omega$ (which can also be found from the unphysical divergence at $\omega_i\rightarrow \Omega$). }\label{fig:profile25}
\end{figure}

In Fig.~\ref{fig:spectra05_B1g}, we plot the total Raman response at $T=0.02$ for the $B_{\textrm{1g}}$ symmetry channel with $U=0.5$ for different energies of the incident photons. This case corresponds to a moderately correlated metal in the high-temperature phase, with a CDW gap of size 0.5. The total Raman response function for the $B_{\textrm{1g}}$ symmetry contains two contributions: the nonresonant contribution (dashed line), which is the only contribution at very high photon energies $\omega_i\to\infty$, and the resonant contribution, which is also the total (resonant) response for the $B_{\textrm{2g}}$ symmetry. For small values of $\omega_i$, we observe only a continuous enhancement of the spectra until $\omega_i$ is large enough to create excitations across the smallest subgaps in the thermally excited DOS. The $\omega_i=0.3$ and $\omega_i=0.346$ curves correspond to the initial transition of the electron from the lower CDW band to the upper subgap states with a further transition to the lower subgap states with an energy loss around $\Omega\sim(U-E)/2=0.16$ and from the lower subgap states to the upper CDW band with a further transition to the upper subgap states with an energy loss around $\Omega\sim E=0.18$ [see panel (a) for details]. In addition, there is a peak which corresponds to the two particle excitations around $\Omega\sim(3E-U)/2=0.02$. When the energy of the incident photons is tuned out resonance with these subgap states (e.g. $\omega_i=0.4$), the intensity of the peaks rapidly decreases until we approach the next resonance at $\omega_i=U$, which corresponds to the initial transitions from the lower to upper CDW bands, with further transitions to all states below. In this case, we observe the largest resonant enhancement [see panel (c)] of more than a factor of 1000. Note that we also have ``joint'' resonance effects, as there are multiple peaks resonanting with this incident photon energy, but the resonance rapidly decreases and becomes small again once $\omega_i$ reaches about 0.7 [see panel (b)]. Increasing the incident photon energy further leads to a continuous decrease of the resonant response without any significant change in its shape; the high energy peak simply moves to the higher frequencies and the response settles into the nonresonant one. Note that every curve shows a large peak in the limit where $\Omega\to\omega_i$.  This peak is an artefact of the infinite-dimensional limit and the hypercubic lattice, and is not expected to be seen in any real material system.

  Similar behaviour is observed for the case of a strongly correlated insulator (in the normal state) at $U=2.5$ and $T=0.06$ in Fig.~\ref{fig:spectra25_B1g}. The main differences with the previous case are connected with two points. First, the gap is larger and the subgap states are wider separated. Hence, the response is very small in the low-energy part of the spectrum and for low initial photon frequencies. Second, the single particle excitation energies are quite different. For the case of $U=0.5$, the energies of the single particle excitations $(U-E)/2=0.16$ and $E=0.18$ are close to each other and the corresponding peaks of the response functions effectively merge. Now these peaks at $(U-E)/2=0.4$ and $E=1.7$ are well separated and can be distinguished in the spectrum. In addition, as was seen for the $A_{\textrm{1g}}$ total Raman response in the normal state of the Falicov-Kimball model~\cite{sh_v_f_d2}, the mixed contribution becomes large and negative for large enough values of the transfered frequency $\Omega$ and can completely cancel the resonant contribution when one is in the Mott insulator phase~\cite{MSF3}. Moreover, for some values of $\Omega$ the sum of the mixed and resonant contributions is negative and the total Raman response for the $A_{\textrm{1g}}$ symmetry becomes smaller than the nonresonant one (not shown here).

Another important feature to examine in the total Raman response is the resonant profile of the response, which is a cut through the spectra with a fixed value of the transferred energy $\Omega$ while varying the incident photon frequency $\omega_i$. In Figs.~\ref{fig:profile05} and \ref{fig:profile25}, we plot the total Raman response functions for different symmetries at various (fixed) transferred frequencies $\Omega$ as a function of the incident photon frequency $\omega_i$.

In the case of $U=0.5$, for small values of the transferred frequency $\Omega=0.1$, we observe a wide peak centered around $\omega_i\sim0.3$, which correspond to the joint resonance when $\omega_i\sim(U+E)/2$ is tuned to the single particle transitions from the lower CDW band to the upper subgap states and $\omega_f=\omega_i-\Omega\sim E$ is tuned to transitions between the lower and upper subgap states. Another sharp peak at $\omega_i=0.6$ corresponds to transitions with the scattered frequency $\omega_f=U$. For larger values of the transferred frequency $\Omega$, the resonant profiles become more complicated and dramatically change as the transferred frequency is increased. This complicated behavior is caused by the requirement to satisfy the resonance conditions when the frequencies $\omega_i$, $\omega_f$, and $\Omega=\omega_i-\omega_f$ must be tuned to the available single particle transitions. Due to this constraint not all of the main resonances are seen, like the one at $\omega_i=U$. But for the large values of the transferred frequency $\Omega\gtrsim U$, when only transitions between the lower and upper CDW bands are involved, the shape of the resonant profiles changes smoothly and slowly.

For large values of $U=2.5$, when the peaks of the single particle DOS (as well as the energies of the single particle transitions) are well separated, the resonant profiles display much more complicated behavior (see Fig.~\ref{fig:profile25}, where we show just one symmetry channel, since all channels are very similar on the log scale). The overall profiles are significantly enhanced when the transferred frequency is larger than about 0.8.  The profiles also illustrate peaks which change shape dramatically as $\Omega$ is changed.  Such behaviour is similar to what was seen for the resonant profiles in the normal state~\cite{sh_v_f_d4}.

\section{Conclusions}

In this work, we have shown how one can solve for the exact total electronic Raman response of a CDW
insulator that is formed via a nesting instability.  Since the DOS reconstructs significantly below $T_c$, we also see a significant change in the Raman response as a function of $T$.  The exact solution is made possible for the Falicov-Kimball model in the infinite-dimensional limit, where DMFT is exact.  We use the Falicov-Kimball model because the charge vertex is known exactly for this model.

Our main results are that there are a large number of strong resonances associated with all of the different peaks in the ordered-phase DOS, which has significant subgap states at low $T$. The strongest resonance occurs between the states separated by $U$ corresponding to the $T=0$ gap.  Since most CDW systems have gaps less than an electron volt, this resonance would not normally be able to be seen with optical light.  If the incident photon frequency is larger than the gap, we can, nevertheless, see some joint resonances, where lower-energy peaks resonate, similar to what was seen in in previous normal state calculations.  In any case, we feel these results indicate that there should be very interesting Raman scattering structures seen in experiment when one examines resonant effects in materials where the ordering yields a divergence in the single-particle DOS at $T=0$, such as the CDW case we examined here. Hopefully, these kinds of experiments will be undertaken soon.

More interesting is the case when the incident photon energy can be on the order of the CDW gap.  To do this, we need to find materials with larger gaps than most currently known CDW systems.  Perhaps these kinds of materials can be found in the future and the experiments we envision carried out on them as well. 

In any case, what is clear is that resonant effects to electronic Raman scattering in ordered systems can yield a wide range of interesting results, even if a microscopic description of the physical behavior is challenging.

\acknowledgments

J.~K.~F.~acknowledges support from the Department of Energy, Office of Basic Energy Science, under Grant No.~DE-FG02-08ER46542.  The collaboration was supported by the Department of Energy, Office of Basic Energy Science, under Grant No.~DE-FG02-08ER46540.

\end{document}